\documentstyle[twoside,fleqn,espcrc2]{article}
\newcommand{\be}{\begin{equation}} \newcommand{\ee}{\end{equation}} \newcommand{\ba}{\begin{eqnarray}}
\newcommand{\ea}{\end{eqnarray}} \newcommand{\la}{\label}

\newcommand{\tr} {\hbox{tr}}

\newcommand{\AmS}{{\protect\the\textfont2
  A\kern-.1667em\lower.5ex\hbox{M}\kern-.125emS}}

\title{Strings and Branes in Nonabelian Gauge Theory\hfill {\normalsize CERN-TH/99-198}}

\author{Christof Schmidhuber
\address{CERN, Theory Division, CH-1211 Gen\`eve 23}
        \thanks{Lecture based on talks given in Berlin, Bern and at
Cern (1998/99). This work is supported in part by the DFG.}
} 
\input epsf
\begin{document}

\begin{abstract}

\noindent
It is an old speculation that SU(N) gauge theory 
can alternatively be formulated as a string theory. 
Recently this subject has been revived, in the
wake of the discovery of D-branes.
In particular, it has been argued that at least some conformally invariant 
cousins of the theory have such a string
representation. This is a pedagogical introduction to
 these developments for
 non-string theorists. Some of the existing arguments are simplified.

\end{abstract}

\maketitle

\section{Introduction}

It is an old speculation, motivated by work of Wilson \cite{wil}, `t Hooft \cite{hoo}, Polyakov \cite{pol}
and many others, that non-abelian gauge theories can alternatively be
formulated as string theories. Recently this subject has been revived, in the
wake of the discovery of Dirichlet-branes \cite{polch} and the investigation of their properties.
In particular, based on work by 
Polyakov \cite{polya} and Klebanov \cite{kle},
Maldacena \cite{mal} has
argued that at least $N=4$ super-Yang-Mills theory has such a string
representation; the argument has been extended in
\cite{kgp}, \cite{wit1} and in many other papers.

The following is a summary of some of these developments,
intended especially for non-string theorists. 
This is mainly a review, although I will use various new arguments
and simplified derivations.
For a more complete
review and an extensive list of references, see \cite{mal.sum}.

We consider $SU(N)$ gauge theory with gauge field $A_\mu^aT_a$, where $T_a$
are the generators of $SU(N)$ in the adjoint representation. The Lagrangean is
$${\cal L}\ =\ {1\over g^2}\int d^4x\ \tr\ {1\over4}F_{\mu\nu}F^{\mu\nu}\ ,$$
where $F_{\mu\nu}=\partial_{[\mu}A_{\nu]}+i[A_\mu,A_\nu]$ is the Yang-Mills field
strength and $g$ is the coupling constant. 
There might also be a certain number $n_F$ of massless flavors in the
fundamental representation of the gauge group.

 \begin{figure}[htb]
 \vspace{9pt}
\vskip3cm
 \epsffile[1 1 0 0]{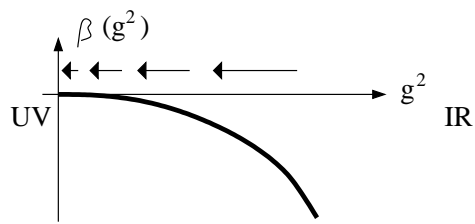}
\caption{Running coupling constant for QCD}
\end{figure}

An important property of the coupling constant $g$ is that it flows under scale 
transformations:
$${d\over d\log\mu} g^2(\mu)\ \ =\ \ \beta[g^2(\mu)]\ ,$$
where $\mu$ is the scale at which, e.g., scattering experiments are performed.
The one-loop-beta function
$$\beta(g^2)\ \sim\ -({11\over3}N-{2\over3}n_F)\ (g^2)^2$$
has a coefficient that is negative as long as the number of flavors is not too big.
Then the theory is asymptotically free, flowing to weak coupling in the UV
and to strong coupling in the IR. More precisely, in the UV
$$g^2(\mu)\ \sim\ {1\over\log{\mu\over\Lambda}}\ ,$$
where $\Lambda$ is the scale at which $g$ becomes of order 1 ($\sim$ 350 MeV
in QCD).

This behavior of the coupling constant means that perturbation theory in $g$
is o.k. at short distances (say, much smaller than the size of a nucleus),
but is useless in the IR where the theory is strongly coupled. But of course
it is often the IR properties of gauge theories that we are particularly
interested in. E.g., we would like to see that QCD accounts for quark confinement.
In the simpler case of the pure gauge theory, we would like to see that
the Wilson loop obeys an area law. 

When at least two flavors are present,
we would like to see whether and how the global chiral symmetry
is spontaneously broken to its diagonal subgroup.
And of course we would like to compute baryon masses and compare them
with experiment, or (already in the pure gauge theory) compute glueball masses.
Also, we would like to study QCD at finite temperature, i.e. under conditions
present in the early universe. Is there a phase transition at some critical
temperature, above which confinement is lost and chiral symmetry is restored?
In order to adress any of these questions, we need methods to study
nonabelian gauge theories at strong coupling.

In this talk I will focus on the case $n_f=0$ without flavors.
Then the coefficient of the quadratic beta function coefficient is
simply proportional to $N$, and it is useful to redefine the coupling
constant to $\lambda=g^2N$, such that $N$ disappears from the flow equation:
$$\dot \lambda\sim -\lambda^2\ \ \ \ \hbox{with}\ \ \ \ \lambda=g^2N\ .$$

\section{Lattice Gauge Theory}

One way of studying strongly coupled {\it euclidean} gauge theory is to regularize
the theory by putting it on  lattice with lattice spacing $a$,
using, e.g., the Wilson lattice action, and to then take the continuum limit.
If the theory is considered at large {\it bare} coupling constant $\lambda_{bare}$,
then the partition function and all correlation functions can be expanded in a power 
series in $\lambda_{bare}^{-1}$. It is well-know that in this expansion e.g. the
partition function can be written as a sum over closed surfaces on the lattice (Fig. 2),
with some surface tension $\sigma$, weighted by the `t Hooft factor
$N^{2-2g}$ where $g$ is the genus of the surface:
$$Z\ \sim\ \sum_{cl.surf.\Sigma}\exp\{-\sigma\cdot Area(\Sigma)\}\ N^{2-2g}\ .$$
In particular, in the ``planar limit''
$$g\rightarrow0\ ,\ \ N\rightarrow\infty, \ \lambda=g^2N\ \hbox{fixed}\ ,$$
only surfaces of spherical topology (genus 0) survive.
The question naturally arises, whether there is a continuum limit of this picture:
if we go to the continuum limit of the gauge theory by taking the lattice
spacing $a$ to zero while adjusting the bare coupling constant $\lambda$ such
that the renormalized coupling constant remains fixed, is then the
partition function of the gauge theory given by a sum over closed
continuous surfaces?
I.e., is SU(N) gauge theory described by a string theory with string tension
$\sigma$ and string coupling constant $\kappa={1\over N^2}$?
In particular, is SU(N) gauge theory in
the large-N limit described by some classical string theory?
 
 \begin{figure}[htb]
 \vspace{9pt}
\vskip3cm
 \epsffile[1 1 0 0]{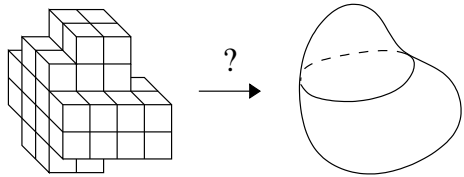}
\caption{Lattice versus continuum gauge theory}
\end{figure}

The same questions can be asked for Wilson loops, i.e. for the trace of the path-ordered
exponential of the gauge field, integrated along a closed contour $C$:
$$W(C)\ =\ <\tr\ P\ e^{i\oint_C A}>\ .$$
In the lattice theory at strong bare coupling, $W(C)$ is given by a sum over
surfaces on the lattice bounded by $C$ (Fig. 3). Is in the continuum theory $W(C)$ 
given by a sum over continuous surfaces with some finite surface tension $\sigma$,
and can this be used to derive an area law for the Wilson loop?

 \begin{figure}[htb]
 \vspace{9pt}
\vskip4cm
 \epsffile[1 1 0 0]{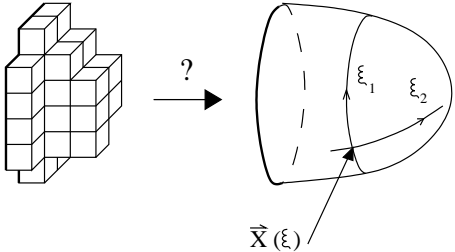}
\caption{Wilson loops}
\end{figure}

There is, of course, a problem. The strong coupling expansion is an expansion
in the inverse {\it bare} coupling constant. But the coupling constant runs.
It runs in such a way that in order to keep the renormalized coupling fixed,
we have to take the bare coupling constant (the coupling constant at scale 
$\mu\sim{1\over a}$) to zero:
$$\lambda_{bare}\ \sim\ {1\over|\log a|}\ .$$
Thus we cannot trust the strong coupling expansion any more, and we cannot
prove from the lattice regularization that the partition function is given
by a sum over surfaces.

One possible way out is to consider instead the ${\cal N}=4$ supersymmetric version
of the theory. One of the nice properties of ${\cal N}=4$ $SU(N)$ super-Yang-Mills
theory is that the coupling constant does not run: $\dot\lambda=0$.
So we can have a continuum theory that is simultaneously strongly coupled
in the IR and in the UV. We will return to this case below and
identify the string theory by which it indeed seems to be described,
following the work mentioned above.
As for the
non-supersymmetric theory, we will say in the end which problem in
two-dimensional conformal field theory needs to be solved in oder to decide 
whether it is a simple string theory or not.

But for the moment let us stick with the non-supersymmetric theory, let us
simply assume that it has a string representation, and let us study what
consequences this would have. That is, we parametrize the surfaces that are
bounded by the loop $C$ by world-sheet coordinates $\xi_1,\xi_2$.
The embedding coordinates of the surface are $X^\mu(\xi)$. So we assume that the Wilson loop
is given by a path integral
\ba W(C) &\sim& \int_{X^\mu|_C=\hat X^\mu(\xi_1)}\ [dX^\mu]\ \times\\
&&\exp\{-\sigma\int d^2\xi
 {\sqrt{\det\ \partial_\alpha X^\mu\partial_\beta X_\mu}}\}\ \ea
Here we have used the Nambu-Goto world-sheet action, which expresses the area
of the world-sheet in embedding space as the integral over
the square root of the determinant of the induced metric on the world-sheet.
In string theory, the tension $\sigma$ is conventionally called ${1\over\alpha'}$.
The boundary values $\hat X^\mu(\xi_1)$ are the embedding coordinates of the loop $C$.

We would now like to review that {\it if} such a string theory of SU(N) gauge theory
exists, it must have at least two perhaps unexpected properties: first,
the string theory must live in five -- rather than four -- dimensions; and second,
it cannot be a bosonic string theory but is probably a superstring theory.

{}\section{D-branes in gauge theory}

Why does the QCD string have to live in more than four dimensions?
Here we follow an argument due to Polyakov \cite{pol.ur}.
The nonpolynomial Nambu-Goto action is difficult to handle in the quantum theory.
As is usual in string theory, we rewrite it with the help of an auxiliary
two-dimensional world-sheet metric $h_{\alpha\beta}$:
\ba &&e^{-\sigma\int d^2\xi{\sqrt{\det \partial_\alpha X^\mu\partial_\beta X_\mu}}}
 \\ &\rightarrow& 
\int [dh_{\alpha\beta}(\xi)]\ e^{-\sigma\int d^2\xi{\sqrt h}h^{\alpha\beta}
\partial_\alpha X^\mu\partial_\beta X_\mu}\ .\la{max}\ea
Classically, $h$ is a non-propagating field.
It is easy to check that the path integral over $h$ has a saddle point
where the auxiliary metric is equal to the world-sheet metric,
\ba h_{\alpha\beta}\ =\ \partial_\alpha X^\mu \partial_\beta X_\mu\ .\la{anna}\ea
The saddle point value of the integrand indeed reproduces the Nambu-Goto action.
The embedding coordinates $X^\mu$ are now free fields coupled to the metric $h$.
We will return to the boundary conditions for $h$ in a moment.  

A two-dimensional metric has three components, and there are two diffeomorphisms.
Thus there is one gauge invariant degree of freedom which can be chosen to be
the conformal factor. I.e., every two-dimensional metric, at least on a surface
with the topology of a disc, can be written as
$$h_{\alpha\beta}\ =\ e^{\phi(\xi)}\hat h_{\alpha\beta}\ ,$$
where $\hat h$ is an arbitrarily chosen background metric on the world-sheet
that nothing physical can depend on.

Now, classically, even the conformal factor $\phi$ drops out of the action
in (\ref{max}).
Quantum mechanically, there is the conformal anomaly:
$${\delta S_{eff}\over\delta\phi(x)}\ \equiv\ {\sqrt h}<T^\mu_\mu>\ \sim\ 
c\ {\sqrt h}R^{(2)}\ =\ -2\Box\phi\ ,$$
where $c$ is the ``central charge''. Integrating this equation shows that
the effective action contains a piece proportional to
$$\int d^2\xi {\sqrt{\hat h}}\hat h^{\alpha\beta}\partial_\alpha\phi\partial_\beta\phi\ .$$
Comparing with (\ref{max}), we see that
the conformal factor enters exactly as if it were another embedding coordinate:
the string effectively lives in
the five-dimensional space $(X^\mu,\phi)$.

 \begin{figure}[htb]
 \vspace{9pt}
\vskip5cm
 \epsffile[1 1 0 0]{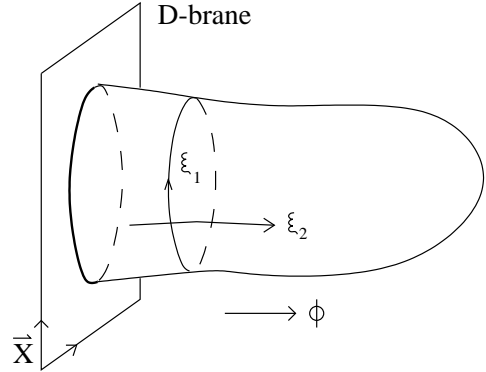}
\caption{The Dirichlet-brane}
\end{figure}

Where in this five-dimensional space is the four-dimensional space in which
the Wilson loop lives? To this end we consider the issue of boundary conditions
for the metric $h_{\alpha\beta}$. In oder to be consistent with the
saddle point equation (\ref{anna}) for $h$, we choose
$$h_{11}|_C\ =\ \partial_1X^\mu \partial_1X_\mu|_C\ .$$
(More precisely, there is a zero-mode corresponding to constant shifts of $\phi$,
which is now also fixed by this boundary condition.)
Now, the background metric $\hat h$ could be arbitrary, but it is convenient
to choose it such that at the boundary:
$$\hat h_{11}|_C\ =\ \partial_1X^\mu \partial_1X_\mu|_C\ .$$
This implies that, at the boundary,
$$\phi|_C\ =\ 0\ .$$
This is a Dirichlet boundary condition.
It means that, although the bulk of the world-sheet can move in five dimensions,
its boundary is restricted to lie on a four-dimensional hyperplane
(see Fig. 4).
In string theory, such $(p+1)$-dimensional hyperplanes on which
otherwise closed strings can end are called
``Dirichlet $p$-branes''. So at best we can hope to recover SU(N) gauge
theory as the world-brane theory that lives on a Dirichlet-3-brane in a 
higher-dimensional string theory. This was the first remark.

{}\section{Fermionic Strings}

The second remark is that the string theory cannot be a bosonic string theory
\cite{pol}.
To see this, let us first consider the simpler Ising model. It consists of
spins $\sigma_i$
whose values can be either +1 or --1,
sitting on  the sites of a lattice.
The energy of a configuration is the sum over all links $<ij>$ of the product
of the neighboring spins $\sigma_i$ and $\sigma_j$. The partition function
is the sum over all spin configurations, weighted with the Boltzmann factor:
$$Z_{Ising}\ =\ \sum_{\{\sigma_i\}}\ \exp\{-\beta\sum_{<ij>}\sigma_i\sigma_j\}\ ,$$
where $\beta$ plays the role of $1/g^2$ in the gauge theory.

Just like the partition function of the gauge theory can be written,
in the strong coupling expansion, as a sum over closed surfaces,
the partition function of the Ising model can be written as a sum over closed
paths:
$$Z_{Ising}\ =\ \sum_{cl. paths}\ e^{-M(\beta)\cdot L}\ ,$$
where $L$ is the total length of the paths and $M=\log(2\tanh \beta)$ 
can be interpreted as the mass of a particle moving on the lattice.
So at first sight it looks as if in the continuum
limit the Ising model was equivalent to a theory of free scalar particles,
whose world-lines are these paths.
 
\begin{figure}[htb]
 \vspace{9pt}
\vskip3cm
 \epsffile[1 1 0 0]{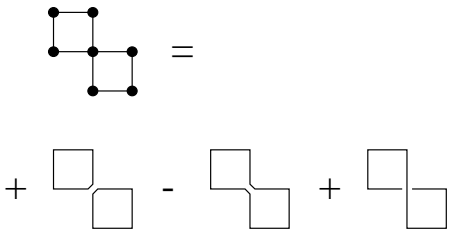}
\caption{Paths in the Ising model}
\end{figure}

But we know that this is not true: instead, the Ising model in the continuum
limit is equivalent to a theory of free fermions. An intuitive way to see this
is the following. Consider the configuration of links drawn in Fig. 5.
There are three ways to represent this lattice configuration in terms of
closed paths, as shown on the right-hand side. So if we describe the Ising
model by free scalar particles, we make a mistake: we count one configuration
three times. We can correct the mistake by weighing each configuration with
a relative factor
$$(-1)^n\ ,$$
where $n$ is the number of 360${}^o$ rotations that the tanget vectors to the paths
make as they go around the paths. This corrects the mistake because it subtracts the
middle path, but now we see that the particle behaves as a fermion: its wave function
flips its sign when the particle is rotated by 360 degrees.

There is no similarly clear argument for the case of surfaces on a lattice,
but it is clear that there are plaquette configurations that we overcount if
we interpret the sum over surfaces on the lattice as a sum over bosonic
string world-sheets. And it is reasonable to expect that we can correct this
mistake by introducing fermions on the string world-sheet. 

One can now
try to guess or derive from first principles just what kind of fermionic string
it is that we need. 
This has in fact been tried for a long time \cite{mig} without clear result.
But there is a simple string theory that contains fermions on the world-sheet
 {\it and} admits Dirichlet-3-branes: this is the type IIB string theory.
Type IIB string theory lives in 10 - not in 5 - dimensions.
But it can be turned into a 5-dimensional string theory
by Kaluza-Klein ``compactification'' on
a compact 5-dimensional manifold $K^5$, such as the 5-sphere $S^5$.
So in the following let us simply 
``try out'' these 5-dimensional superstring theories:
that is, instead of starting with a gauge theory and trying to
construct a dual string theory out of it,
we start with these string theories
 and try to see what kind of
gauge theories on the 3-brane they describe.

To admit it right away: following Maldacena \cite{mal}, we will get not the
standard $SU(N)$ Yang-Mills theory but - depending on what
the compactification manifold $K^5$ is -
various of its 
conformally invariant cousins, including the $N=4$ supersymmetric theory.
So the target will be missed, but not by too much, and we will mention in
the end 
how one might be able to get to realistic gauge theories.

{}\section{The open bosonic string}

The purpose of the next three sections is to explain that
the picture in figure 4 is  oversimplified in the following respect.
It might seem as if the 3-brane
was just a fictitious object that sits in flat, empty 5-dimensional embedding
space without affecting it. This is not true. First,
it turns out that the 3-brane is electrically charged, in a sense that I 
will explain; in the case of $SU(N)$ gauge theory, the charge is proportional
to $N$. And second,
the 5--dimensional space around the D--brane is {\it curved}.
It turns out to be anti-de Sitter space $AdS_5$.

In order to explain these points, we need to spend a few minutes reviewing
some basic facts of string theory.
To keep things simple, it is useful to first consider the open bosonic
string without D-branes (Fig. 6). That is, the boundary of the string world-sheet
is allowed to fluctuate all over the embedding space, which in the case
of the bosonic string is 26-dimensional.

 \begin{figure}[htb]
 \vspace{9pt}
\vskip3cm
 \epsffile[1 1 0 0]{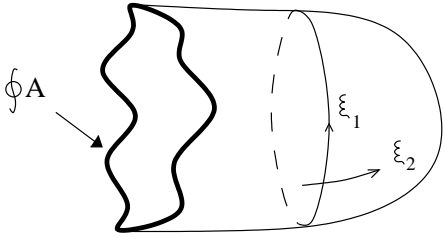}
\caption{The Open Bosonic String}
\end{figure}

We consider the two-dimensional field theory that lives on the string world-sheet.
There are 26 world-sheet fields $X^m(\xi)$, all of which obey Neumann 
boundary conditions.
The world-sheet action is
\ba&&{1\over\alpha'}\int d^2\xi\ \partial_\alpha X^m\partial^\alpha X^n g_{mn}(X)
\la{hugo}
\\ &+& {1\over{\sqrt{\alpha'}}}\oint\partial_\alpha X^m A_m(X) d\xi^\alpha\ .
\la{axel}\ea
$g_{mn}(X)$ is the embedding space metric, which -- from the point of view of
the world-sheet theory -- represents the possibility of adding
generally complicated interactions between the fields $X^m$
(there is also a tachyon, a dilaton and an antisymmetric tensor field
which we suppress for now).
The boundary interactions of the world-sheet theory are similarly
represented in embedding
space by the 26-dimensional gauge field $A_m(X)$. 

A consistency condition in classical string theory is that the world-sheet
theory must be conformally invariant. One way to understand this is as follows:
In section 3, we wrote the world-sheet metric as $h_{\alpha\beta}=\hat h_{\alpha\beta} \ e^\phi$.
$\phi$ became one of the embedding coordinates. $\hat h$ was chosen arbitrarily,
and we have already said that nothing physical can depend on this choice. 
In particular, the world-sheet theory must be completely invariant
under rescaling the background metric $\hat h$.

Scale invariance means that all the beta functions in 
the theory must vanish. If $A_m=0$, the beta function(al) for the
metric $g_{mn}$ in (\ref{hugo}) is well--known to be to lowest order 
in $\alpha'$:
$${d\over d\log\mu}\ g_{mn}\ \sim\ R_{mn}\ ,$$
where $R_{mn}$ is the Ricci tensor of $g_{mn}$.
So we get the Einstein equations (in vacuum). More generally, the
$\beta=0$ conditions of $2d$ field theory are, by definition, the string
equations of motion. 
The string effective action is defined as the action whose 
variation yields these equations of motion. In the low--energy
limit, it is given by \cite{ftc}:
$$ S\ =\  S_{Gravity}\ +\ S_{Gauge}\ +\ o(\alpha')$$
with
\ba S_{Gravity}&\sim&\int d^{26}x\ {\sqrt g}\ {1\over\kappa^2}\ R^{(26)}[g]\\
S_{Gauge}&\sim&\int d^{26}x\ {\sqrt g}\ {1\over\kappa}\ F_{mn} F^{mn}\ \ ,\ea
where $\kappa$ is the string coupling constant,
$R^{(26)}$ is the Ricci scalar and $F_{mn}$ is the field strength
of the gauge field $A_m$. We see that
the gauge coupling constant is related to the string coupling constant by
\ba g^2_{YM}\ =\ \kappa\ .\la{cocos}\ea
The above action is, to lowest oder in $\alpha'$, the action of Einstein gravity coupled to Maxwell theory.
The $\alpha'$ corrections denote the difference between Einstein
 gravity + Maxwell
theory and {\it exact} conformal invariance of the world-sheet theory. 

In the above,
 we have considered world-sheets with the topology of a disc. We can also
allow topology fluctuations in the bulk or at the boundary
of the world--sheet (Fig. 7). Those
correspond to string loop corrections -- i.e. loop corrections to the
gravity action and the gauge action, respectively. Formally,
the classical action must then be replaced by the full effective action.
For the bosonic string,
this is of course not well-defined but for the superstring it is.

 \begin{figure}[htb]
 \vspace{9pt}
\vskip3.5cm
 \epsffile[1 1 0 0]{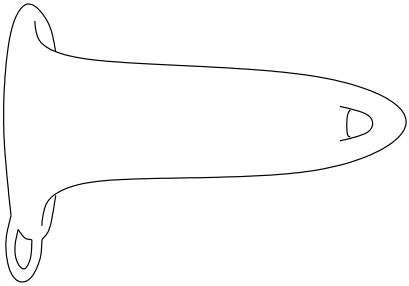}
\caption{Two types of loop corrections}
\label{fig:largenenough}
\end{figure}

For the moment, though, let us stay with classical string theory
and now restrict 
 the boundary of the world-sheet  to lie on a 
Dirichlet 3-brane. This is done
 by switching the boundary conditions
for all but four of the world--sheet fields $X^m$
from ``Neumann'' to ``Dirichlet''. Then 
the Maxwell theory is also restricted to live on
the 3-brane. $S_{Gauge}$ becomes a 4-dimensional action,
delta-function restricted to the brane:
\ba S &=& S^{(26)}_{Gravity}\ +\ S^{(4)}_{Gauge}\times\delta^{22}(x)\la{mira}\ea
The fields that $S^{(4)}_{Gauge}$ contains are
 the 26 components of the gauge field
$A_m$ as before, but now they split up into the
 4 components parallel to the brane
which make up a 4-dimensional Maxwell field $A_\mu$, plus
the 22 components transverse to the brane.
The latter just become scalar fields that live on the brane.

So we learn two interesting things from bosonic string theory: 
first of all, the embedding space metric
cannot be anything. It must obey - to lowest order in $\alpha'$ - 
 the Einstein equations derived from the action (\ref{mira}). 
Away from the brane, this implies
$$R_{mn}=0 \ .$$

Second, there is a dynamical gauge field $A_\mu$
that lives on the 3-brane. This helps answering a question: in
the first part of the talk we have started with a
4-dimensional gauge theory
and looked for its ``dual'' string theory. Given such a string theory,
how can we recover the original gauge field?
We would like to identify it with the 3-brane gauge field we have
just found.
However, so far this is only an abelian gauge field.
To get a nonabelian gauge field we must consider a generalization:
Instead of a single D-brane we formally consider a set of $N$ D-branes
sitting on top of each other (figure 8). The Wilson loop now carries
a ``color'' index running from 1 to $N$ (labelling the D-brane on
which it ends), and one can argue that the
Maxwell field is replaced by a SU(N) gauge field \cite{wit.N}.

\section{Superstrings}

How is the discussion
modified if we consider type IIB superstring theory
instead of bosonic string theory? One modification is quite expected:
the gauge theory that lives on the brane also becomes supersymmetric.
The other crucial difference between D--branes in bosonic string theory
and D--branes
in superstring theory is that the latter are charged.
Let us begin with the first modification.

In the case of the superstring, there are only 10 (and not 26)
embedding coordinates $X^m$. 
Viewed as world--sheet fields, four of them 
(say $X^\mu$ with $\mu\in\{0,1,2,3\}$)
obey Neumann boundary conditions, 
and six of them (denoted by $X^t, t\in\{4,...,9\}$)
obey Dirichlet boundary conditions. We will denote by $X^4$
the radial coordinate of this transverse space, and by $X^5,...,X^9$
its angles.

If the world-sheet boundary could lie anywhere
in the embedding space 
(i.e. if all coordinates had Neumann boundary conditions),
the gauge theory would be 10-dimensional super-Yang Mills theory which contains
the gauge field $A_m$ and a
gaugino $\psi$ with 16 spinor components.
But since the Yang--Mills theory  is restricted to the 3-brane,
$A_m$ splits up into a 4-dimensional gauge field $A_\mu$ plus
six scalars $A_i$, while the 10-dimensional gaugino splits up into four 
4-dimensional gauginos $\psi^a$ (since
4-dimensional gauginos have only four spinor components).
This results in precisely the field content
of ${\cal N}=4$ supersymmetric Yang-Mills theory. 
So this is why one expects a dual gauge theory with up to 4 supersymmetries.
So much for the first modification.

Next, let us discuss in what sense
the 3--branes are ``charged''. In the case of the bosonic string,
the bulk theory was 26--dimensional
bosonic gravity. In the case of type IIB superstring theory,
the bulk theory is what is called 
``type IIB supergravity'' \cite{jhs} in the 10-dimensional embedding space.
Apart from the metric $g_{mn}$, type IIB supergravity contains a variety of fields.
But the only one that
 we need to mention here is
the so-called self-dual Ramond-Ramond 4-form gauge field $C^{(4)}$.
This is an antisymmetric tensor field with 5-form 
field strength $F^{(5)}$. The gauge invariance consists
of adding to $C^{(4)}$ the total derivative of a 3--form 
$\Lambda^{(3)}$. ``Self-duality'' means that
 $F^{(5)}$ is equal to its dual $*F^{(5)}$:
\ba 
F^{(5)} &=& d\wedge C^{(4)}\\
\delta C^{(4)} &=& d\ \Lambda^{(3)}\\
F^{(5)}  &=& *F^{(5)}
\ea
The 10--dimensional ``gauge field'' $C^{(4)}$ must of course
 not be confused with the 1-form $A_\mu$
that lives only on the 3--brane.

It is this 
4-form gauge field $C^{(4)}$, with respect to which  
-- as realized by Polchinski \cite{polch} --
the 3--brane is charged: 3--branes are
sources of ``electric field strength'' $F_{01234}$ (again, ``4''
denotes the radial direction of the transverse
space).
Each 3-brane carries
one unit of Ramond-Ramond charge:
$$\oint_{S^5} *F^{(5)}\ \sim\ N\ $$
where the 5--sphere $S^5$ surrounds the 3--brane in 10 dimensions.
Moreover, because of self-duality, there is also a component $F_{56789}$:
 each brane also carries
one unit of magnetic charge:
$$\oint_{S^5} F^{(5)}\ \sim\ N\ .$$
Here it is useful to think of a D-brane in 10--dimensional
 type IIB superstring theory as an an analog of
a charged particle in ordinary 4--dimensional Maxwell theory.
In this analogy, $F^{(5)}$ in 10 dimensions is
replaced by the Maxwell field strength $F^{(2)}$ in 4 dimensions,
and the $S^5$ is replaced by an $S^2$ surrounding the particle.
The D--brane is then analogous to a dyon, carrying simultaneously 
electric and magnetic charge.

Let us conclude our brief review of
the relevant facts of string theory with the
10--dimensional bulk action that replaces $S_{Gravity}^{(26)}$
in (\ref{mira}). It contains the terms
\ba S^{(10)}&\sim& \int d^{10}x\ {\sqrt g}\ {1\over\kappa^2}\  
R^{(10)}[g]\\ &+& \int d^{10}x\ {\sqrt g}\ 
 {1\over2\cdot5!}\ F_{mnpqr}F^{mnpqr}\ .\ea
As a result, the Einstein equations away from the brane
 change to
\ba
R_{mn}\ \sim\ \kappa^2\ F_{mabcd}F_n^{\ abcd}\ .\la{ruth}
\ea
In other words, the electric flux $F^{(5)}$ emanating from the branes
carries energy-momentum and therefore curves the 5-dimensional
space-time. Let us next discuss what the curved space looks like.

{}\section{Anti-de Sitter space}

From now on we will
consider the 10-dimen-sional string theory, ``compactified''
down to five dimensions. These five dimensions will now be denoted
by $x^m$ with $m\in\{0,1,2,3,\phi\}$. The remaining coordinates
$x^i$ with $i\in\{5,6,7,8,9\}$ parametrize the compact manifold
$K^5$, whose size and shape we take to be constant, i.e.
independent of the $x^m$.
All fields $g_{mn},C^{(4)}$ are assumed to be independent
of the compact coordinates $x^i$. The 3-branes are
again parametrized by $x^\mu,\ \mu\in\{0,1,2,3\}$ and are assumed to be
``smeared out'' over $K^5$.

 \begin{figure}[htb]
 \vspace{9pt}
\vskip4.5cm
 \epsffile[1 1 0 0]{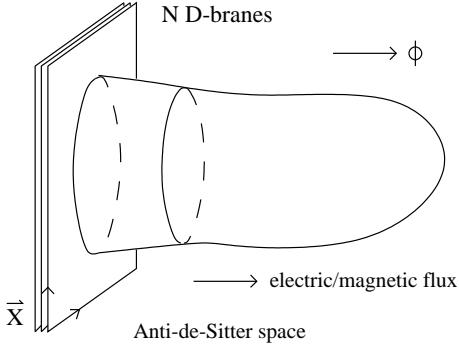}
\caption{Charged D-branes}
\end{figure}

The 3--branes are then analogous to charged capacitor plates:
the electric charge of the 4--dimensional
 branes simply creates a constant density of
electric flux $F_{0123\phi}$ in the 5-dimensional space-time.
This density is proportional to $N$, 
because there are $N$ branes.
It is also inversely proportional to the Volume Vol${}_K$
of the 5-dimen-sional compactification space $K^5$, because
-- upon replacing the $S^5$ by $K^5$ in the previous section --
the integral of $F^{(5)}$ over $K^5$ gives one unit of charge:
$${F_{0123\phi}\over {\sqrt{g^{(5)}}}}\ \propto\ {N\over \hbox{Vol}_K}\ .$$
Here, $g^{(5)}$ is the determinant of the 5-dimensional metric  $g_{mn}$.
From the Einstein equation (\ref{ruth}) we see that the curved space
is anti-de Sitter space, $R_{mn}\propto -g_{mn}$: it has constant negative curvature scalar
$$R\ \ \sim\ \ \kappa^2\ F_{0123\phi}F^{0123\phi}\ \ \sim\ \ -\ {\kappa^2N^2\over
\hbox{Vol}_K^2}\ .$$
To determine the volume of $K^5$, we write the metric on $K^5$ as
$$ds_K^2\ =\ L^2\ d\vec\theta^2\ ,$$
where $\vec\theta$ are angular coordinates on $K^5$ (e.g.
$d\vec\theta^2=d\Omega_5^2$ for $K^5=S^5$). Because the
5--form $F^{(5)}$ is self--dual,
the Einstein equation (\ref{ruth})
tells us that $K^5$ is an Einstein manifold, $R_{ij}\propto +g_{ij}$, whose
Ricci scalar 
$$R_K\ \sim\ {\kappa^2N^2\over
\hbox{Vol}_K^2}\ $$
has the opposite sign but the same magnitude 
as that of the anti-de Sitter space-time.
Since the Ricci scalar is $\sim L^{-2}$
and Vol${}_K\sim L^5$ we conclude
that the curvature
radius is
\ba {L\over\sqrt{\alpha'}} \sim\  (\kappa N)^{{1\over4}}\ .\la{yy}\ea
Here we have reinstated $\alpha'$ to make $L$ dimensionless.

To summarize, the $N$ branes carry ``Ramond-Ramond charge'',
and the resulting flux curves the five-dimensional space,
thereby turning it into anti-de Sitter space.
One way of writing the metric on anti-de Sitter space is
\ba ds^2\ =\ d\phi^2\ +\ \exp\{-{2\over L}(\phi-\phi_0)\}\ dx_{||}^2\ .
\la{moritz}\ea
Here, $x_{||}$ denotes the coordinates $x^\mu$ parallel to the D-branes,
and we are assuming that the branes are to the left.
Note however that there is a free parameter $\phi_0$; it will be given an
interpretation later.

We have derived this result to lowest order in $\alpha'$
and for a surface with disc topology.
Let us discuss corrections to this result \cite{kle,mal}.
We have already mentioned three types of corrections.
First, there are corrections in $\alpha'$. Those come out to be proportional to
$$\alpha' R^{(10)}\ \ \sim\ \ {\alpha'\over L^2}\ \ \sim\ \ {1\over\sqrt{\kappa N}}
\ \ \sim\ \ {1\over\sqrt{\lambda}}\ ,$$
where $\lambda=g^2N=\kappa N$ is the coupling constant of the super-Yang-Mills theory, using (\ref{cocos},\ref{yy}).
So the anti-de Sitter space solution
that we have found in the supergravity approximation to superstring theory
can be trusted (away from the brane)
as long as the super-Yang-Mills coupling constant $\lambda$
is strong.
As $\lambda$ is decreased, one has to replace the supergravity equations
of motion by the conditions of exact conformal invariance of the world-sheet
theory.

The second type of correction comes from
topology fluctuations on the world-sheet boundary, as in figure 7.
Those correspond to loop corrections to the super-Yang-Mills theory,
which are power series in $\lambda$ and therefore large.
As a result, the classical action $S_{Gauge}^{(4)}$ is replaced
by the full quantum effective action $\Gamma_4$. 
So our supergravity solution contains information about
this effective action in the limit of large $\lambda$ 
-- precisely the limit we were interested in! 
On the other hand, the supergravity solution is {\it not} appropriate
in the perturbative regime of small $\lambda$.

The third type of corrections is due to
topology fluctuations in the bulk of the world-sheet.
Those correspond to loop corrections to supergravity. They
come out to be proportional to
$$\kappa^2(\alpha'R^{(10)})^4\ \sim\ {\kappa^2\alpha'^4\over L^8}\ \sim\ {1\over N^2}\ .$$
This is precisely what one expects from a string theory that describes
a gauge theory: non-planar diagrams are suppressed by
powers of ${1\over N^2}$! In particular, the limit $N\rightarrow\infty$
can be taken, in which the 
 {\it classical} supergravity solution can be trusted
and string loop corrections can be neglected.

{}\section{Scales and Wilson loops}

We would like to conlude this review with
the project that is suggested by figure 3:
to compute a Wilson loop by computing the minimal area
of the string world-sheet which it bounds.
What we have learned in the meantime is
that this world-sheet lives in a 5-dimensional curved space
with anti-de Sitter metric (\ref{moritz}), while its boundary
(the Wilson loop) lives on
a flat four-dimensional hypersurface located at $\phi=0$.

Before  the Wilson loop can be computed, 
the free parameter $\phi_0$ in the metric (\ref{moritz})
must be interpreted, because
 obviously the result will depend on it.
The interpretation
is that $\phi_0$ defines the scale in the gauge theory in the following
 sense:
a scale transformation
\ba x_{||} &\rightarrow& e^\tau x_{||}\la{one}\ea
of the 4-dimensional physical
space parametrized by $x_{||}$ is equivalent in (\ref{moritz})
to a shift of $\phi_0$:
\ba\phi_0&\rightarrow&\phi_0+L\tau\ .\la{two}\ea
It is useful to redefine
$$ z\ =\ {L\ e^{\phi-\phi_0\over L}}\ ,$$
so that the metric becomes independent of the free parameter $\phi_0$:
\ba\ ds^2\ =\ {L^2\over z^2}(dz^2 + dx_{||}^2)\ \ 
.\la{lucy}\ea
We see that anti-de Sitter space has a boundary at $z=0$.
$\phi_0$ now enters in terms of the location of the branes.
The branes sit at $\phi=0$, i.e. at
$$z\ =\ z_0\ \equiv\ Le^{-{\phi_0\over L}}\ .$$
From (\ref{one},\ref{two}) we then see that
translating the D-branes in $z$ corresponds
to probing different scales in the four-dimensional gauge theory.
And in particular, going to the ultraviolet limit $\phi_0\rightarrow\infty$ 
of the gauge theory corresponds to
placing the branes at the boundary $z=0$
of anti-de Sitter space (compare with \cite{suss}).
In the following computation, $z_0$ will be regarded as an ultraviolet
cutoff. The gauge theory on the brane is
then the bare gauge theory, while properties of the renormalized gauge 
theory will be read off from the interior of anti-de Sitter space.
As $z_0$ is taken to zero, the bare parameters will be adjusted
to keep these renormalized parameters fixed.

So let us illustrate these remarks at the example
of the Wilson loop in the large-$N$, large-$\lambda$ limit,
outlining a calculation by Maldacena and by Rey and Yee \cite{mal.W}. 

 \begin{figure}[htb]
 \vspace{9pt}
\vskip4.5cm
 \epsffile[1 1 0 0]{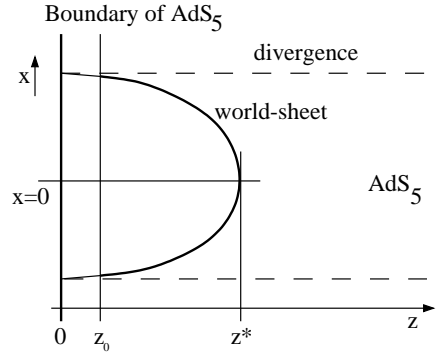}
\caption{Computing Wilson loops}
\end{figure}

In Minkowski space, it is convenient to let the ``loop''
consist of two lines of length $T$ 
parallel to the time axis, separated by a spatial distance $l$.
We define $l$ to be 
measured in the metric $dx_{||}^2$ (rather than ${L^2\over z^2}dx_{||}^2$).
A constant time slice through the world-sheet is 
a line $z(x)$ in the $x-z-$plane, as drawn in
figure 9. We denote by $z^*$ the maximal value of $z$,
$$z^*\ =\ z_{max}\ ,$$
and pick the origin of the $x$--axis such that $x=0$ for $z=z^*$.
The Nambu-Goto action (the area of the world-sheet) becomes $T\cdot E$, where
\ba E\ =\ \int dx\ {L^2\over z^2}{\sqrt{1+(z'(x))^2  }}\ \la{pepa}\ea
is the energy of a pair of oppositely charged particles, separated by 
the distance 
$$l\ =\ 2x(z_0)\ .$$
The equations of motion of this action can be seen to 
yield the following minimal area world-sheet:
\ba |x(z)|\ =\ z^*\int^1_{z/z^*}{d\zeta\over {\sqrt{{1\over\zeta^4}-1}
}}\ .\la{pepe}\ea

Let us now define the continuum limit as follows:
as discussed above, we think of the gauge theory on the brane
as the bare gauge theory with ultraviolet cutoff $z_0$.
We take this cutoff $z_0$ to zero, 
thus placing the branes at the boundary of anti-de Sitter space.
Keeping the renormalized physics fixed means keeping $z^*$ fixed.
This requires adjusting $l$, which has a finite limit:
$$l(z_0)\ =\ 2x(z_0)\ \rightarrow {(2\pi)^{3\over2}\over\Gamma({1\over4})^2}\ 
z^*\ \ \ \hbox{as}\ z_0\ \rightarrow\ 0\ $$
(On top of this, $l$ has a dimension, but this has already been
taken care of by measuring $l$ using the metric $dx_{||}^2$.)

Next, we can compute the area of the minimal area world-sheet
by differentiating (\ref{pepe}) and plugging the result into (\ref{pepa}).
In the limit $z_0\rightarrow0$,
the area  diverges because
$$x(z)\sim z^3\ +\ \hbox{const.}\ \ \ \hbox{near}\ \ z=0\ 
\ \  \rightarrow\ \ E\sim{1\over z_0}\ .$$
The divergence is equal to
the length of the two dashed lines the figure. 
Its interpretation is the following:
The energy $E$ consists of two contributions: one from the electrostatic
attraction of the
charged particles, and the other one from their mass.
In computing the Wilson loop, the latter must be subtracted; it is equal to
the energy of
massive {\it uncharged} particles. The dashed lines just represent 
the world--sheets of such uncharged particles.
Subtracting this divergence and performing the integral, 
one finds \cite{mal.W}: 
\ba E 
&=& -\ {4{\sqrt2}\pi^2\over\Gamma({1\over4})^4}\ {\sqrt{\lambda}\over l}\ .\la{miri} \ea
What do we learn from this result? First,
we see that the energy obeys a Coulomb
law, rather than being linear in $l$ as would 
have been expected in a confining theory.
Indeed, the  ${\cal N}=4$ theory is conformally invariant and not confining.

So the Wilson loop does {\it not} obeys an area law.
In the above calculation this arises essentially because 
the world-sheet drops towards the center of AdS so quickly that
the bulk of the world-sheet gives a vanishing contribution to
its area. A confining theory would have to be described by a
different 5--dimensional geometry (not $AdS_5$), in which
the world--sheet cannot drop to the center (see \cite{wit2}).

What is unexpected in (\ref{miri}) is the
factor ${\sqrt\lambda}$: perturbatively
one would have expected a factor of $\lambda=g^2N$.
So this is an example of a nontrivial claim of the string representation
of non-perturbative
N=4 supersymmetric gauge theory \cite{mal}.

{}\section{Outlook}

To conclude,
at least for the conformally invariant {\cal N}=4 supersymmetric
version of large--N Yang--Mills theory
we now seem to have a strong-coupling description in terms of
a particular 5-dimensional string theory, expanded around
an $AdS_5$ background. Previously, 
explicit string representations for Yang--Mills theories have been
known only in the two--dimensional case \cite{gro}. 

Much work has followed this ``discovery'', of which I will mention
just a few examples.
First, one can also compute the correlation functions of local operators,
following Gubser, Klebanov, Polyakov
\cite{kgp}
and Witten \cite{wit1}.
This has been used to predict the scaling dimensions of various operators 
in the large-$N$, large-$\lambda$ limit \cite{kgp,wit1}.

Second, replacing the compactification manifold $S^5$ by a more general
Einstein manifold $K^5$, the dual string theories have been identified
as certain exotic conformally invariant, but not always supersymmetric
four-dimensional gauge theories \cite{klewit}. To identify them,
it helps to
compare global symmetries. E.g., the 
$SO(6)$ symmetry of the $S^5$ calls for being identified
with the SU(4) $\sim$ SO(6) R-symmetry 
of the ${\cal N}=4$ super-Yang-Mills theory that 
rotates the four gauginos into each other \cite{mal}.
All these theories must be conformal, as the $SO(2,3)$ symmetry of $AdS_5$
reflects the $SO(2,3)$ conformal group in four dimensions.
So there appears to be a rich set of RG fixed points and flows between them
in four-dimensional gauge theory. The author's own work in this
context can be found in \cite{schmi}.

 \begin{figure}[htb]
 \vspace{9pt}
\vskip5cm
 \epsffile[1 1 0 0]{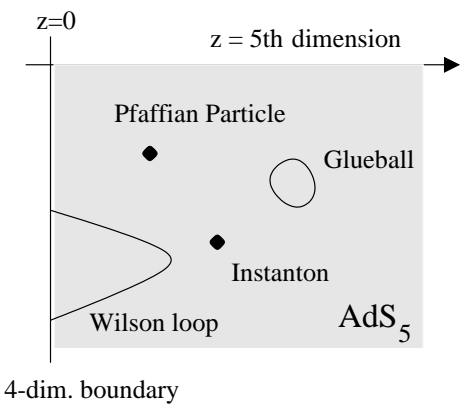}
\caption{Objects in strongly coupled {\cal N}=4 supersymmetric gauge theory}
\end{figure}

Third,
it has been pointed out
that the 5-dimensional AdS-space
is in fact
 populated with all kinds of elementary and solitonic objects.
E.g., instantons in the Yang-Mills theory have been identified
with instantons of the 5-dimensional string theory, with
their position in
the fifth dimension just representing the scale modulus of the instanton
solution \cite{gree}. Furthermore, it is an old conjecture that baryons appear as solitons
in the QCD string theory. In our case there is no chiral matter, and
there are no baryons - but in the case of $SO(N)$ gauge theory
there are analogs of baryons, the ``Pfaffian particles''.
Those have indeed been identified with string theory solitons -- 
namely five-branes,
wrapped over the $K^5$ \cite{wit.bary}.
These and other objects are symbolically drawn in figure 11.

So far we have been discussing the ${\cal N}=4$ supersymmetric
or other conformally invariant versions
of SU(N) gauge theory. But what about the
standard, asymptotically free non-supersymmetric Yang-Mills theory
we started with? Can we draw
a similar picture as  in figure 10?

There is a suggestion due to Witten how to generalize the
string representation to non-supersymmetric gauge theory
 in principle \cite{wit2}.
Using this prescription, some qualitative features such as confinement
have been argued to emerge as expected \cite{gro.oog}.
There is even a quantitative
computation of glueball masses (see e.g. \cite{mal.sum}).
However, the problem is that 
this defines a theory with strong bare coupling $\lambda$.
As mentioned in the beginning, what we need in the non-supersymmetric
case is weak bare coupling and strong renormalized coupling.
Just as in lattice gauge theory
at strong bare coupling, there is no reason to expect the results
for glueball masses, etc.,
to be universal \cite{gro.oog}. Universal results are only expected
in the continuum limit
$$a\rightarrow 0\ \ ,\ \ \ \lambda_{bare}\sim{1\over|\log a|}\ .$$
Can this limit be studied in the string theory representation? Since 
$\lambda\rightarrow0$,  the supergravity solution can no longer be trusted.
Instead one has to sum up all 
the $\alpha'$-corrections. In other words, one needs an exact two-dimensional
conformal field theory. 

Many exact conformal field theories in two dimensions
are known, but none of them includes the `Ramond-Ramond' background $F^{(5)}$
that we need here.
So here is the problem that needs to be solved before one can decide whether
bosonic SU(N) gauge theory is ``dual'' to a simple string theory:
extend the current algebra methods of conformal field theory to 
the case of a self-dual Ramond-Ramond background! 

Whether this can be done or whether it is as intractable as the perturbative
approach to QCD remains to be seen.

\newpage

\noindent
{\bf NOTE:}
\vskip 3mm

\noindent
One aspect in which this
review deviates from the standard treatment of the subject
should be mentioned. 
Usually, the 3-branes are introduced in 10 dimensions. 
Then one goes to the near--horizon limit 
of the resulting geometry, which is $AdS_5\times S^5$.
Here we basically first ``compactify''  on the $S^5$
and then indroduce D--branes in 5 dimensions (section 7).
Then there is no need to go to a near-horizon limit:
the $5d$ geometry is automatically $AdS_5$ everywhere.
The embarassment of having to explain 
{\it six} new coordinates (instead of only one) also disappears.
Moreover, the branes are now ``inside the geometry'' before the
continuum limit of the gauge theory is taken.
The continuum limit automatically moves them to the ``correct'' end
of $AdS_5$ -- the boundary (section 8).

\end{document}